\documentclass[]{aa}
\usepackage{graphicx}

\begin{document}
\input{psfig.sty}

\title{Extragalactic Globular Clusters in the Near Infrared:\\
III. NGC~5846 and NGC~7192 
\thanks{Based on observations at the Very Large Telescope of the 
European Southern Observatory, Chile (Program 63.N-0287)} 
\thanks{Based on observations made with the NASA/ESA Hubble Space
 Telescope, obtained from the data archive at the Space Telescope
 Science Institute. STScI is operated by the association of
 Universities for Research in Astronomy, Inc. under the NASA contract
 NAS 5-26555.}}

\subtitle {Quantifying the age distribution of sub-populations}
\author {Maren Hempel
        \inst{1}
        \and
        Michael Hilker
        \inst{2}
        \and
        Markus Kissler-Patig
        \inst{1}
        \and
        Thomas H. Puzia
        \inst{3}
        Dante Minniti
        \inst{4}
        \and
        Paul Goudfrooij
        \inst{5} }

\offprints {Maren Hempel} 

\institute{European Southern Observatory, Karl-Schwarzschild-Str.~2,
85748 Garching, Germany \\ \email{mhempel@eso.org, mkissler@eso.org}
\and
Sternwarte der Universit\"at Bonn, Auf dem H\"ugel 71, 53121 Bonn,
Germany\\
\email{mhilker@astro.uni-bonn.de}
\and
Sternwarte der Ludwigs-Maximilians-Universit\"at, Scheinerstr.~1,
81679 M\"unchen, Germany\\
\email{puzia@usm.uni-muenchen.de} 
\and
Departamento de Astronom\'\i a y Astrof\'\i sica, P.~Universidad
Cat\'olica, Casilla 306, Santiago 22, Chile \\
\email{dante@astro.puc.cl}
\and
Space Telescope Science Institute, Baltimore, USA \\
\email{goudfroo@stsci.edu}  }

\date{Received 10 February 2003 ; accepted 23 April 2003 }

\abstract{
In this third paper of our series on near-IR and optical photometry of
globular cluster systems in early-type galaxies we concentrate on the
photometric results for NGC~5846 and NGC~7192, two group ellipticals,
and on a first comparison between the globular cluster systems
investigated so far. In NGC~5846 the colour-colour diagram shows clear
bi-modality in $(V-K)$, which is confirmed by a KMM test. The mean
colour of both peaks were estimated to be $(V-K)$$_{\rm
blue}$=2.57$\pm0.06$ and $(V-K)$$_{\rm red}$=3.18$\pm0.06$. The
situation in NGC~7192 is different, in that the colour-colour diagram
gives no evidence for a distinct second population of globular
clusters. Using simulated colour distributions of globular cluster
systems, we make a first step in quantifying the cumulative age
distribution in globular cluster systems. Also here the result for
NGC~5846 leads us to the conclusion that its metal-rich globular
cluster population contains two globular cluster populations which
differ in age by several Gyr. The age structure for NGC~7192 shows
instead strong similarity with a single-age
population.}
\authorrunning{Hempel et al.}
\titlerunning{Extragalactic Globular Clusters in the Near Infrared {\tt
III}}
\maketitle

\section{Introduction}
\label{s:intro}

As shown by many studies during the last decade (Zepf \& Ashman~1993,
Ashman \& Zepf~1998, Kundu \& Whitmore~2001a,~b, Larsen et al.~2001,
Kissler-Patig et al.~2002) globular cluster systems are a very
powerful tool in galaxy formation and evolution studies. Although
there is a wide agreement about the existence of sub-populations in
cluster systems regarding their metallicity and their age, the
discussion about the origin of those populations is still
ongoing. Starting with Peebles \& Dicke (1968) who assumed globular
clusters to be the first objects formed in the early universe, the
number of possible formation scenarios increased drastically since
then. One of the main issues is to explain the multi-modality found in
colour-colour diagrams of globular cluster systems. In general it is
agreed that different globular cluster populations are produced during
strong star formation events, but the nature of these events remains
under debate. Besides the merger scenario, favoured by Ashman \& Zepf
(1993) (see also \cite{ash92}, Whitmore et al.~1993; Whitmore \& Schweizer~1995;
Schweizer et al.~1996; Kissler-Patig~2000), there is a number of
alternatives, i.e.~the accretion scenario (\cite{cote98}; \cite{cote02}; \cite{hilker99})
or the monolithic collapse (\cite{forb97}, \cite{kiss98}, and references
therein). From the photometric point of view it has been shown that 
the multi-modality of the colour distribution is a common feature of
globular cluster systems (\cite{gebh99}). In particular, the existence
of a blue, old and metal-poor cluster population (\cite{ash93};
\cite{burg01}) is a 
general feature. Regarding the red sub-population the almost only
consensus which has been reached so far, is about the large varieties
between different cluster systems. This includes the possible
existence of sub-populations within the red population which is
interpreted as a sign of different star formation events in later
stages of the galaxy evolution. 

The main problem in specifying and dating different star formation
events arises from the age-metallicity degeneracy of the 
most commonly used optical colours. So far 
the bulk of high-quality photometric investigations have been performed
using {\it Hubble Space Telescope (HST)\/} in the optical wavelength regime
(\cite{forb98}; \cite{kund98}; \cite{gebh99}; \cite{larsen01}).   
It has been shown (\cite{minniti96}; \cite{kiss00}; \cite{kiss02a}
(hereafter cited as Paper I); \cite{puzi02} (hereafter Paper II)) 
that combining optical and near-infrared data is a more promising
method to separate age and metallicity effects and to access the
relative ages of the globular cluster populations. The method relies on a 
sampling effect, where the V-band is dominated by 
stars near the turn-off (TO) region, whereas the main contribution to the
K-band is from giant branch stars 
(\cite{yi01}). Whereas the TO is dominated by age effects, the giant
branch shows a high sensitivity to metallicity
(\cite{savi00b}). This results in a similar dependence of $(V-I)$ and
$(V-K)$ on the age of the clusters, but a higher sensitivity of
$(V-K)$ to the metallicity. 

In the previous two papers of this series (Papers I and II), a
systematic survey of globular cluster systems of E and S0 galaxies in
the combined optical and near-infrared wavelength range (using the
$V$, $I$, and $K_{\rm s}$ bands) has been started.
In Paper I, we compared the globular cluster systems of two ellipticals in
the Virgo cluster, namely the giant central galaxy M87 and an
intermediate-luminosity galaxy NGC~4478. We found that in those cases, the
$(V\!-\!K)$ colour distribution yielded roughly consistent conclusions
relative to those derived from the $(V\!-\!I)$ colours measured by {\it
  HST}. In Paper II however, the $(V\!-\!K)$ colours of globular clusters in
NGC~4365 led us to postulate the existence of a significant population of
intermediate-age ($\sim$\,2-6 Gyr old), metal-rich globular clusters, which
was not revealed by the $(V\!-\!I)$ colours. This important result has
recently been confirmed by deep spectroscopy (Larsen et al.\ 2003), adding
credibility to the results derived from our optical + near-infrared imaging
program. 

One of the aims of this series is to study the the globular cluster systems
of galaxies in different environments, e.g. giant ellipticals in centres of
clusters as well as rather isolated and less luminous galaxies. The present
work will focus on NGC~5846, a giant E0 galaxy in the centre of the
Virgo-Libra Cloud, 
and on NGC~7192, an isolated elliptical with only one companion. Basic
informations on both galaxies are provided in Table 1. 

Both systems have already been studied in the optical (\cite{forbes97b}; 
\cite{gebh99}), and show a very broad ($V-I$) colour distribution. This work
aims at 
the detection of different globular cluster sub-populations, the
determination of their age structure and a first comparison of various
globular cluster systems. The present paper is organised as
follows. In \S 2 the observations and the data reduction procedures
are described. Chapter 3 contains the main results of the observations
and \S4 describes our approach towards 
quantifying the age structure in globular
cluster systems and the results for both systems. In chapter 5 we will
give an outlook on the upcoming work.

\begin{table}
\centering
\caption[width=\textwidth]{General information about the host galaxies
NGC~5846 and NGC~7192. The references are (1):\cite{deva91}, (2):
\cite{schleg98}, (3):\cite{buta95}, (4): \cite{frog78},
(5):\cite{tonr01}}
\label{tab:galprop}
\begin{tabular}{l r r l}
\hline
\noalign{\smallskip}
Property & NGC5846 & NGC~7192 & Reference\\
\noalign{\smallskip}
\hline
\noalign{\smallskip}
RA(J2000)           & 15h 06m 29s            & 22h 06m 50s            &
(1)\\
DEC(J2000)          & $+01^{\rm o}$ 36' 25'' & $-64^{\rm o}$ 18' 57'' &
(1)\\
$B_{\rm T,0}$       & 10.87                  & 12.19                  &
(1)\\
E$_{B-V}$           & 0.055                  & 0.034                  &
(2)\\
$(B-V)_{\rm o}$     & 0.96                   & 0.92                   &
(1)\\
$(V-I)_{\rm eff,o}$ & 1.28$\pm0.01$          & 1.24$\pm0.01$          &
(3)\\
$(V-K)_{\rm eff,o}$ & 3.51$\pm0.01$          &                        &
(4)\\
$(m-M)_V$           & 31.98$\pm0.20$     & 32.89$\pm0.32$             &
(5)\\
M$_V$               & $-22.07\pm0.20$    & $-21.62\pm0.35$           
&(1),(5)\\
\noalign{\smallskip}
\hline
\end{tabular}
\end{table}


\section{Observations and Data Reduction}
\begin{figure*}[!]
\centering
\includegraphics[width=8cm]{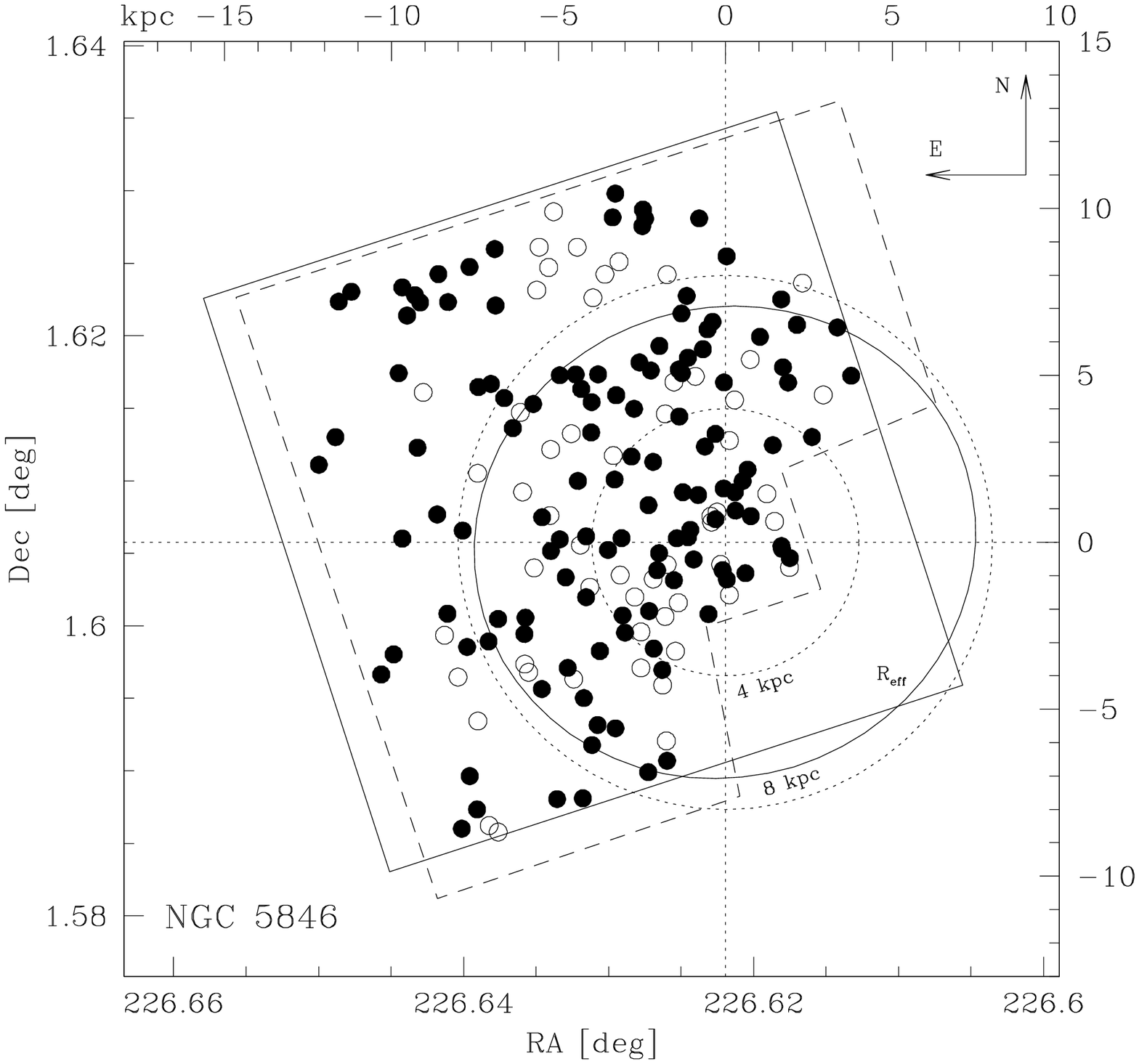}
\includegraphics[width=8cm]{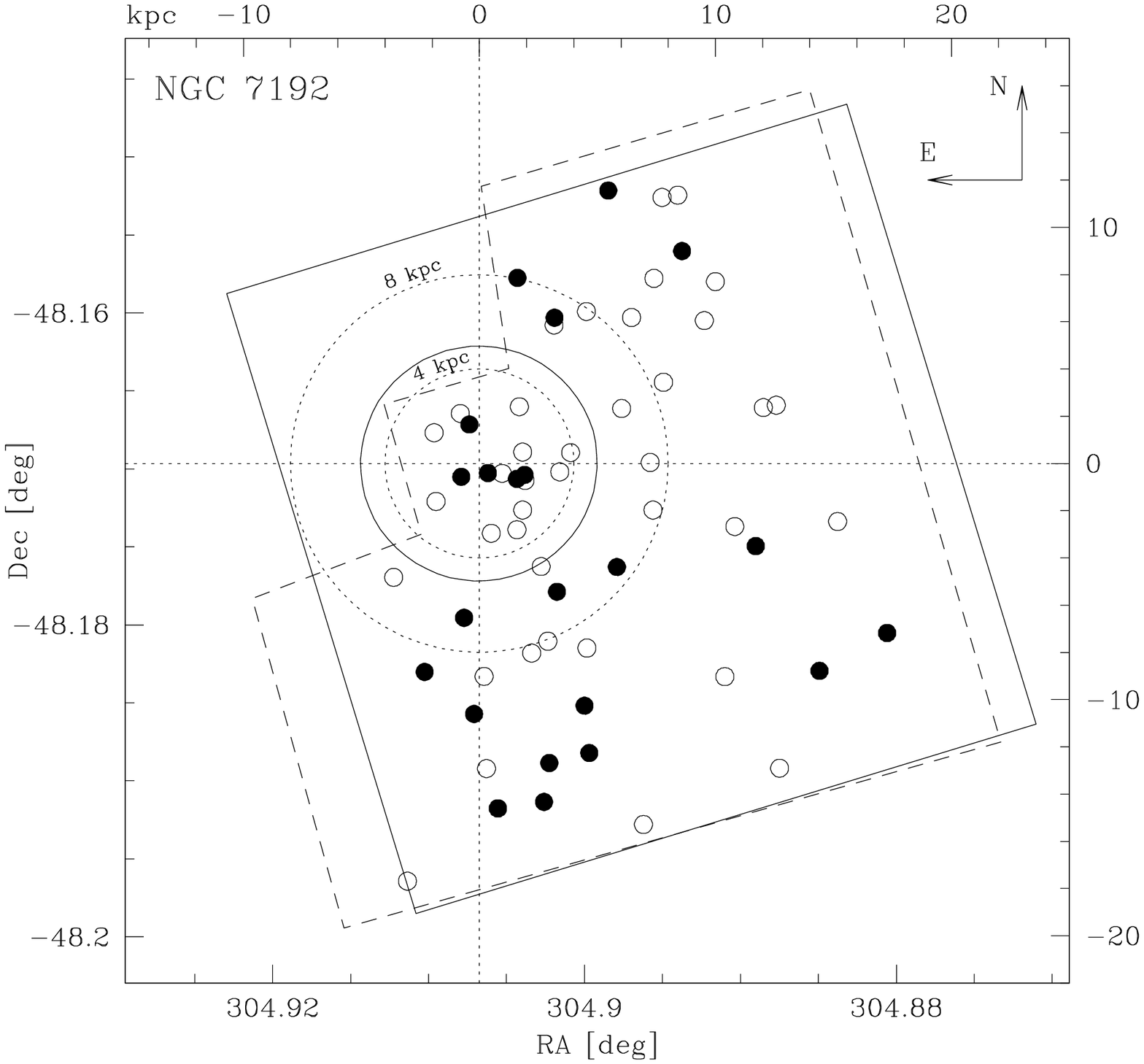}
\caption{Field of view of NGC~5846 (upper) and NGC~7192 (lower).
 The ISAAC field was chosen to fit the available archive data taken
 with WFPC2 onboard the HST. The dashed lines in both frames indicate the
 HST field and the solid one the corresponding ISAAC image. Dotted
 circles show the 4 and 8 kpc distance to the galactic centers, the
 solid circles the half light radii of the galaxy's light. The
 globular cluster samples were divided in two colour
 sub-populations. Dots mark objects with 0.8$\leq$$(V-I)$$\leq$1.2
 and open circles represent clusters in the range 1.2$<$$(V-I$)$\leq$1.5. }
\label{f:fov}
\end{figure*}

\subsection{VLT/ISAAC Near--Infrared Data}

The $K_{\rm s}$ band exposures for both galaxies have been taken in service
mode (ESO program 63.N--0287) with the Near-Infrared Spectrometer And
Array Camera (ISAAC) attached to the Unit Telescope 1 (Antu) of the
European Southern Observatory's Very Large Telescope (VLT). The
field-of-view of ISAAC's Rockwell infrared array is 2.5\arcmin $\times$
2.5\arcmin, with a pixel scale of 0.147\arcsec /pixel. All data were
obtained in April and June 1999 in different nights under varying
photometric and seeing conditions. Standard star observations revealed
that the conditions were not photometric during all nights and
an adjustment (described below) was necessary to align all nights onto
a photometric system. As in Papers I and II we will always refer to the 
$K_s$ filter as $K$.

The NGC~5846 data were obtained during the nights of April 6th, 8th,
and 9th 1999 while the NGC~7192 data were taken during the nights of
June 1st, 2nd, 21st, and 22nd. The observing strategy for NGC~5846
data was the following: 10$\times$10sec object + 5x(2x10sec) sky +
10x10sec object.  The one for NGC~7192: 10$\times$10sec object +
2x(6x10sec) sky + 10x10sec object. The monitoring of the seeing,
magnitude variations, sky level and geometric correction was
done by measuring the parameters of four isolated stars in each single
frame for both galaxies.  A detailed description of the sky
subtraction and combination procedure of the $K$ band data is given in
Puzia et al. (2002). The FWHM of the stellar PSF in the final $K$-band
image is $\sim$0.4\arcsec\ for NGC~5846 and $\sim$0.5\arcsec\ for
NGC~7192.  The total exposure times for NGC~5846 and NGC~7192 are 10000 sec, and
12000 sec, respectively.

\subsubsection{Photometry}
\label{s:calib}
The photometric calibration of the NGC~5846 and NGC~7192 data set was
based on the photometry of 3 and 4 near-IR standard stars, respectively 
(\cite{pers98}). 
 
Measuring the instrumental magnitudes in an aperture of 6
pixel diameter and applying the same analysis as described in Puzia et
al. (2002), the following calibration relations for the photometric
nights have been derived
\begin{eqnarray}
      K_{\mathrm{5846}} & = & k_{\mathrm{inst}} + 23.81(\pm0.018) -
      0.05(\pm0.009)\chi \\
      K_{\mathrm{7192}} & = & k_{\mathrm{inst}} + 23.83(\pm0.010) -
      0.05(\pm0.007)\chi
\end{eqnarray}
where $K_{\rm galaxy}$ is the calibrated magnitude, $k_{\rm inst}$ is
the instrumental magnitude, and $\chi$ the effective airmass ($=$ 1.23
for NGC~5846 and 1.34 for NGC~7192). The error of the zero points
(second term in eq.(1) and (2)) includes photometric errors of each
single standard star measurement and the errors of the aperture correction
analysis. The error of the airmass term is an estimate from the
variations in airmass of all single exposures.

The zero point shifts of the non-photometric nights to photometric
conditions has been derived by tracing the magnitudes of four isolated
stars over all nights. The final photometry was
performed on the overall combined image. For NGC~7192 it was first done on
two combined images separately. One for the photometric nights (June
1st and 2nd), and one for the non-photometric nights (June 21st and
22nd). The magnitudes of both images have been averaged after a
zero point correction of the latter. The true photometric uncertainty,
measured by the scatter of the single measurements is of the order of
$\leq$0.04 mag, mainly due to the strongly varying sky background.

Finally, all magnitudes were corrected for Galactic foreground reddening
using the reddening values of Table 1 and the extinction
curves of Cardelli et al. (1989). The corrections for NGC~5846 and NGC~7192
are $A_K=0.020$ mag and $A_K=0.012$ mag.

\subsection{HST/WFPC2 Optical Data}

The HST data were taken from the public HST archive. NGC~5846 was
observed with HST + WFPC2 under program GO.5920. The total exposure
times of the combined images are 6600 sec in F555W, and 6900 sec in
F814W. NGC~7192 has been imaged with WFPC2 under program GO.5943 in
F555W and F814W filters with 1300 sec and 1000 sec of total exposure
time, respectively. The HST images were reduced and calibrated
following the procedure as described in Puzia et al. (1999, 2002).
All magnitudes were measured with the SExtractor tool using a 8 and
4-pixel-diameter aperture for the PC and WF chip and corrected with
respect to the Holtzman 0.5$\arcsec$ standard aperture
(\cite{holt95}). Instrumental magnitudes were then transformed to the
Johnson $V$ and $I$ magnitudes according to the prescription given by
Holtzman et al. (1995). All magnitudes were reddening corrected using
the following values: $A_V=0.182$ and $A_I=0.107$ for NGC~5846 and
$A_V=0.113$ and $A_I=0.066$ for NGC~7192 (see $E_{B-V}$ in Table 1).


\subsection{Selection criteria}
\label{s:select}
After combining the optical and near-infrared data using the GEOMAP
task within IRAF, the globular cluster sample includes 184 and 61
objects for NGC~5846 and NGC~7192, respectively. Figure
\ref{f:fov} shows the field of view for both galaxies and the spatial
distribution of the cluster candidates. In order to limit the
contamination of the sample by background galaxies or foreground stars
and to set a limit onto the photometric error, general selection
criteria have been applied. In our discussion only objects with an
photometric error $\Delta (V-I)\leq$ 0.1~mag and $\Delta (V-K)\leq$
0.1~mag and a FWHM of the PSF below 0.25$\arcsec$ in the V and I-band are
considered.  

\section{Colour-colour diagrams for NGC~5846 and NGC~7192}
\label{s:colors}

Colour-colour diagrams together with various SSP models (e.g.~in this paper
by Bruzual \& Charlot 2000) are the basis for age and
metallicity estimates. Different models, at identical colors, can show 
differences in {\it absolute} age of about 3 Gyr.
Thus, the SSP approach can only lead to approximate {\it absolute} ages for 
the sub-populations. However, {\it relative} ages can be estimated
accurately enough to separate sub-populations built up during major star 
formation events. 

As shown in Puzia et al.(2002), the various models differ mainly in the
metal-rich range (approx.~[Fe/H]$>-0.8$). 
A detailed comparison between the different models in an absolute sense can 
be found in Maraston et al.~(2001a). 
The metal-rich regime, however, is exactly the one we intend to probe, 
given that the intermediate-age populations are expected to be enriched
in metals. 

As opposed to the situation for absolute ages, the {\it relative} age 
predictions for given colors are relatively similar from model to model. 
Since we focus on the {\it relative}
age dating of globular cluster systems the choice of a specific model is not crucial
and we use the model of Bruzual \& Charlot (2000) throughout the
following analysis. 

As we will show in the following two subsections the results of the
KMM test (\cite{mclachlan88}, updated version 2001) and the
derived metallicities are found to differ significantly from expected
values and even seem to be in contradiction to what visual inspection
of the colour-colour diagrams tells us. Considering the relatively
small sample size, the photometric errors in $(V-K)$ and the limited
depth of the observations (shifting the peak position toward red
colors) the results of the KMM test, as described by McLachlan \&
Basford (1988), have to be discussed with much caution. Since this
work concentrates on relative ages at this stage we will only mention
the results of the KMM and leave the discussion for later.

\subsection{NGC~5846}

\begin{figure*}[!]
\centering
\includegraphics[bb=50 165 580 700, width=12cm]{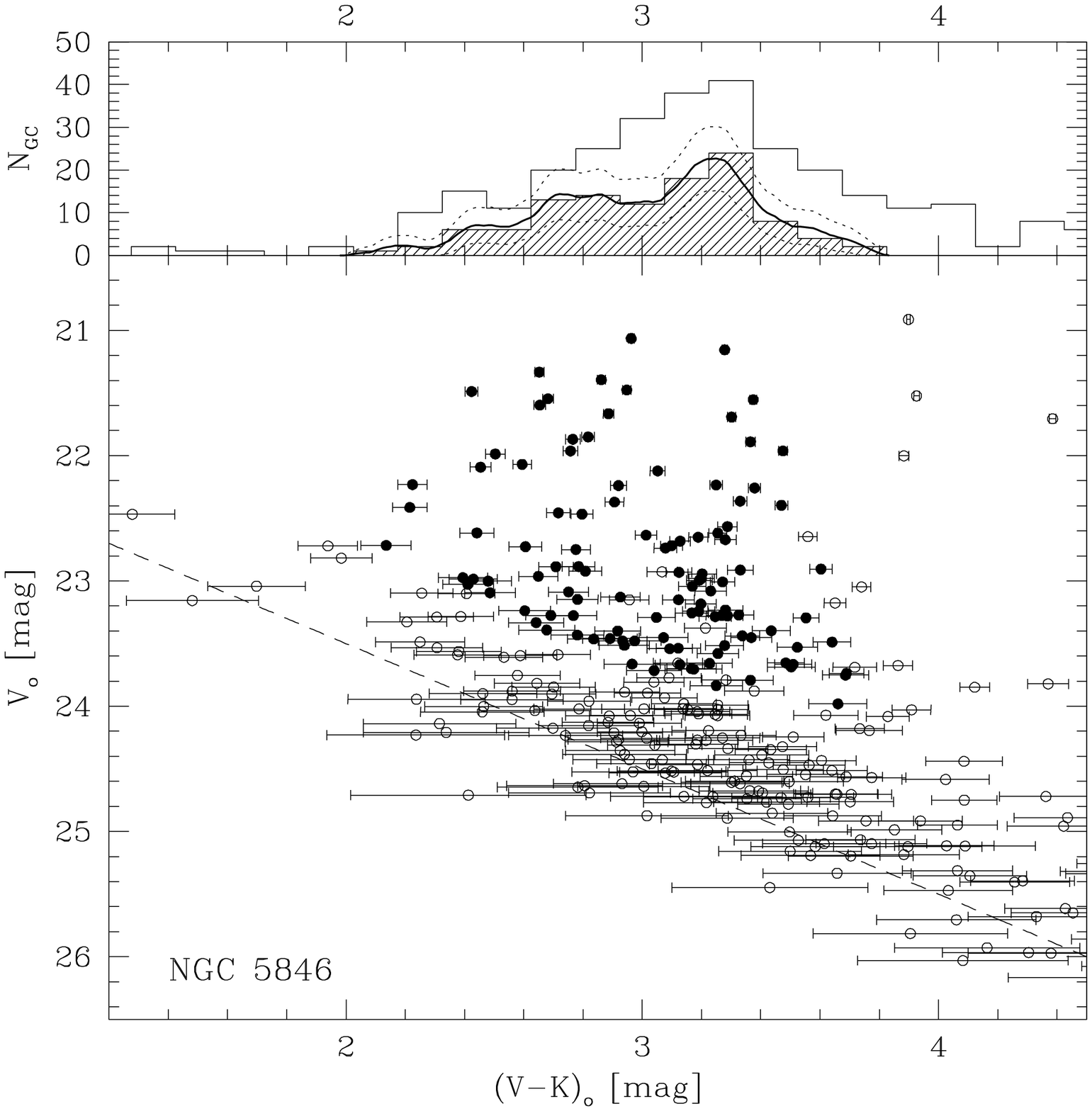}
\caption{$V$ vs.~$(V-K)$ colour-magnitude diagram for NGC~5846.
The top sub-panel shows the colour distribution of 
all (open histogram) and selected (hashed histogram) objects.
The solid line marks the probability density distribution 
together with its 1$\sigma$ uncertainty (dotted line). The lower sub-panel
shows the CMD. Here the
 filled symbols mark the selected clusters while open circles indicate
 rejected objects (see Section \ref{s:select}). Since the photometric
 errors are dominated by the $K$-band, only the $(V-K)$ errors are
 shown. The dashed line marks the limiting magnitude in the $K$-band
($K$ = 21.5 mag).}
\label{f:cmd5846}

\centering
\includegraphics[bb=50 165 580 700, width=12cm]{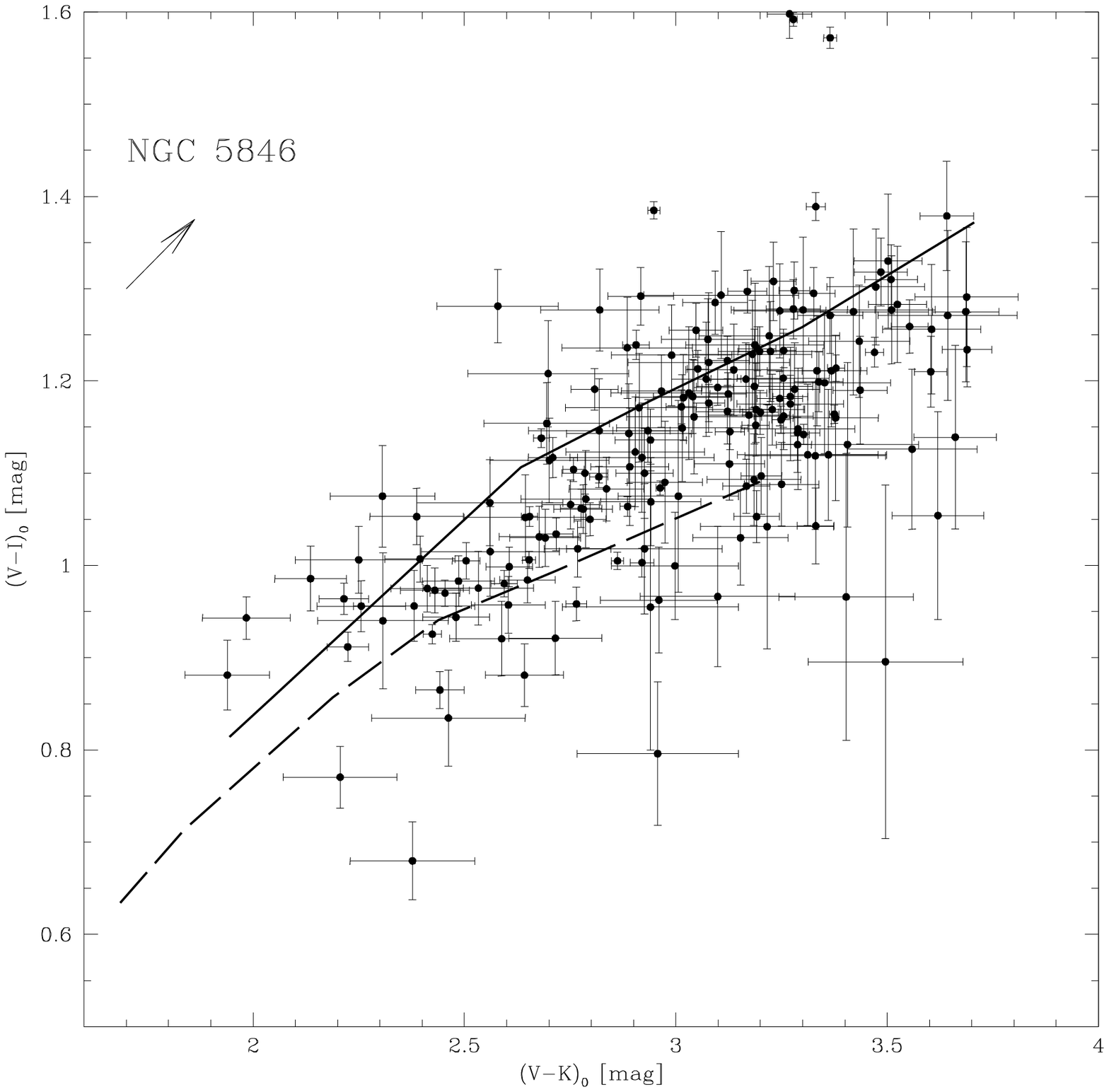}
\caption{$(V-I)$ vs.~$(V-K)$ colour-colour diagram for NGC~5846. All data
are corrected for Galactic foreground reddening. The reddening vector
is marked by the arrow. As example the 15 Gyr and 2 Gyr isochrones (Bruzual \&
Charlot 2000) are marked by a solid and dashed line, respectively. As
for NGC~4365 (\cite{puzi02}) we find a second population of objects
which are red in $(V-K)$ but intermediate in $(V-I)$ with
$(V-I)~\in~[1.0, 1.2]$, and thus are clearly not compatible with old
SSPs.}
\label{f:colcoln5846}
\end{figure*}

The colour-magnitude diagram (hereafter CMD) and the colour-colour diagrams
for NGC~5846 are given in Figures~\ref{f:cmd5846} and
\ref{f:colcoln5846}. The histogram in the upper 
part of the CMD shows the complete set of clusters as a solid histogram
whereas the selected objects, following the selection criteria given
in Section \ref{s:select}, are shown by the hatched histogram.
As expected the selection criteria affect mostly the objects in the
very red colour range, since the photometric errors are larger and
background galaxies possibly contaminate the sample.  

The CMD (Fig.~\ref{f:cmd5846}) reveals evidence for the a bimodal
colour distribution. The KMM test as described by Ashman et al.~(1994)
confirms this result with a confidence level of $\sim$90$\%$ for a
bimodal colour distribution. We obtain peak positions of $(V-K)_{\rm
blue}$ = 2.57$\pm$0.06 and $(V-K)_{\rm red}$ = 3.18$\pm$0.06 for both
colour populations with an estimate of correct allocation value
(confidence level) of 0.79 and 0.94 for the blue and the red peak,
respectively. Hereby about 30$\%$ of the globular clusters were
assigned to the blue population and 70$\%$ to the red. Using the
calibration by Kissler-Patig (Paper I) for old populations, this
corresponds to a metallicity of [Fe/H]=$-0.54\pm0.6$ dex and
$0.27\pm0.4$ dex for the metal-poor and the metal-rich populations,
respectively.

Note that our completeness limit is dominated by the $K$-band, and that we
are biased in favour of red clusters (Paper II; see Fig.~\ref{f:cmd5846}). 
Thus, both the ratio of the numbers of red to blue clusters and the 
metallicity of the blue peak are overestimates.

It is interesting to mention that no multi-modality could be
found in the $V-I$ colour distribution. This is similar to the situation in
NGC 4365 where the intermediate-age population, when projected on the
$V-I$ axis, fills the gap between the two old (metal-rich and metal-poor)
sub-populations. When forced to bimodality, the
$(V-I)$ peak positions determined by the KMM test are
$(V-I)_{\rm blue}=1.11$ and $(V-I)_{\rm red}=1.13\pm0.02$. This is in
agreement with the results by Gebhardt \& Kissler-Patig (1999). \\

\subsection{NGC~7192} 

\begin{figure*}[!]
\centering
\includegraphics[bb=50 150 580 700, width=12cm]{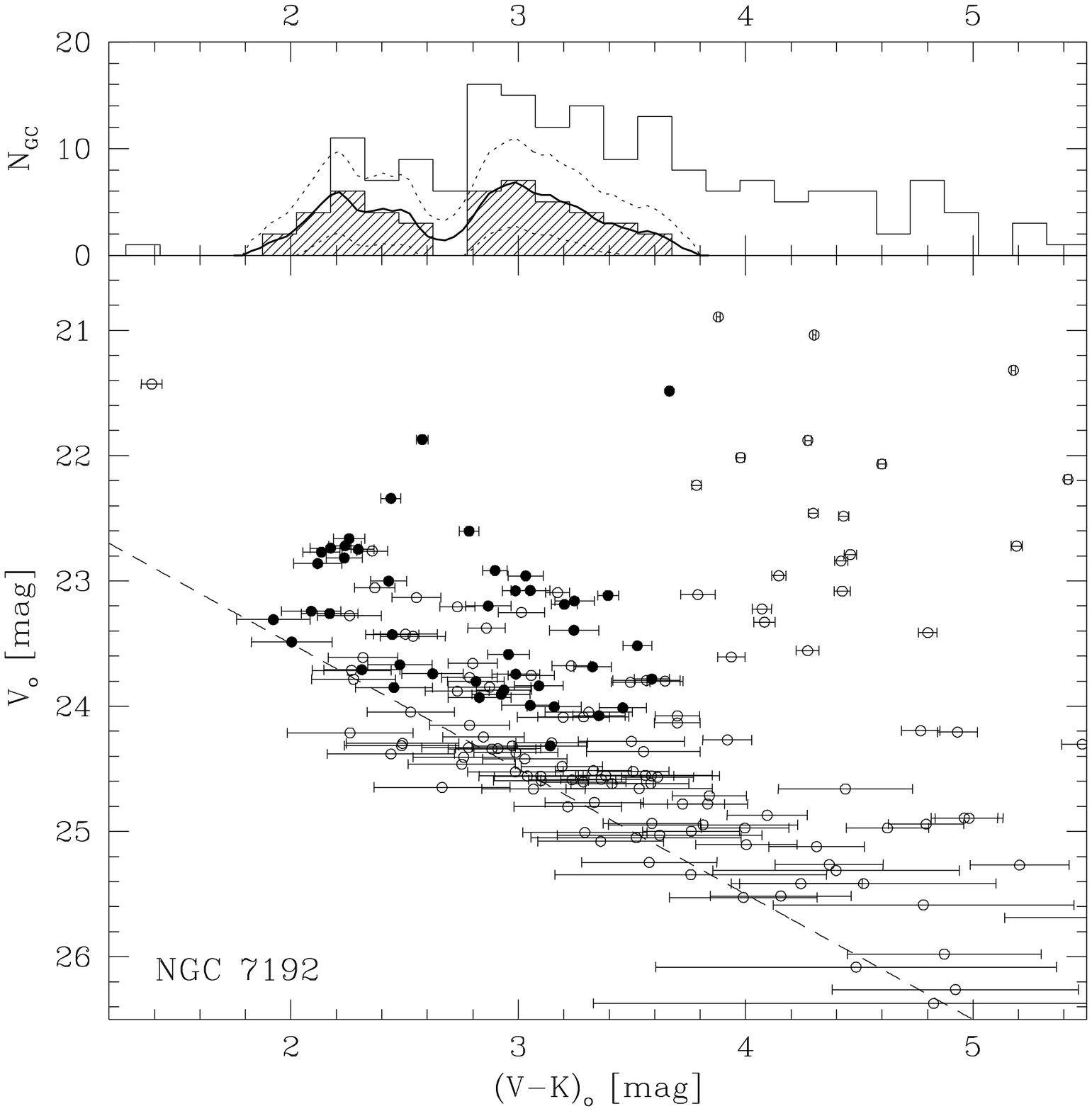}
\caption{$V$ vs.~$(V-K)$ colour magnitude diagram for NGC~7192. 
(panels and symbols as in Fig.~\ref{f:cmd5846}).}
\label{f:n7192cmd}

\centering
\includegraphics[bb=50 150 580 700, width=12cm]{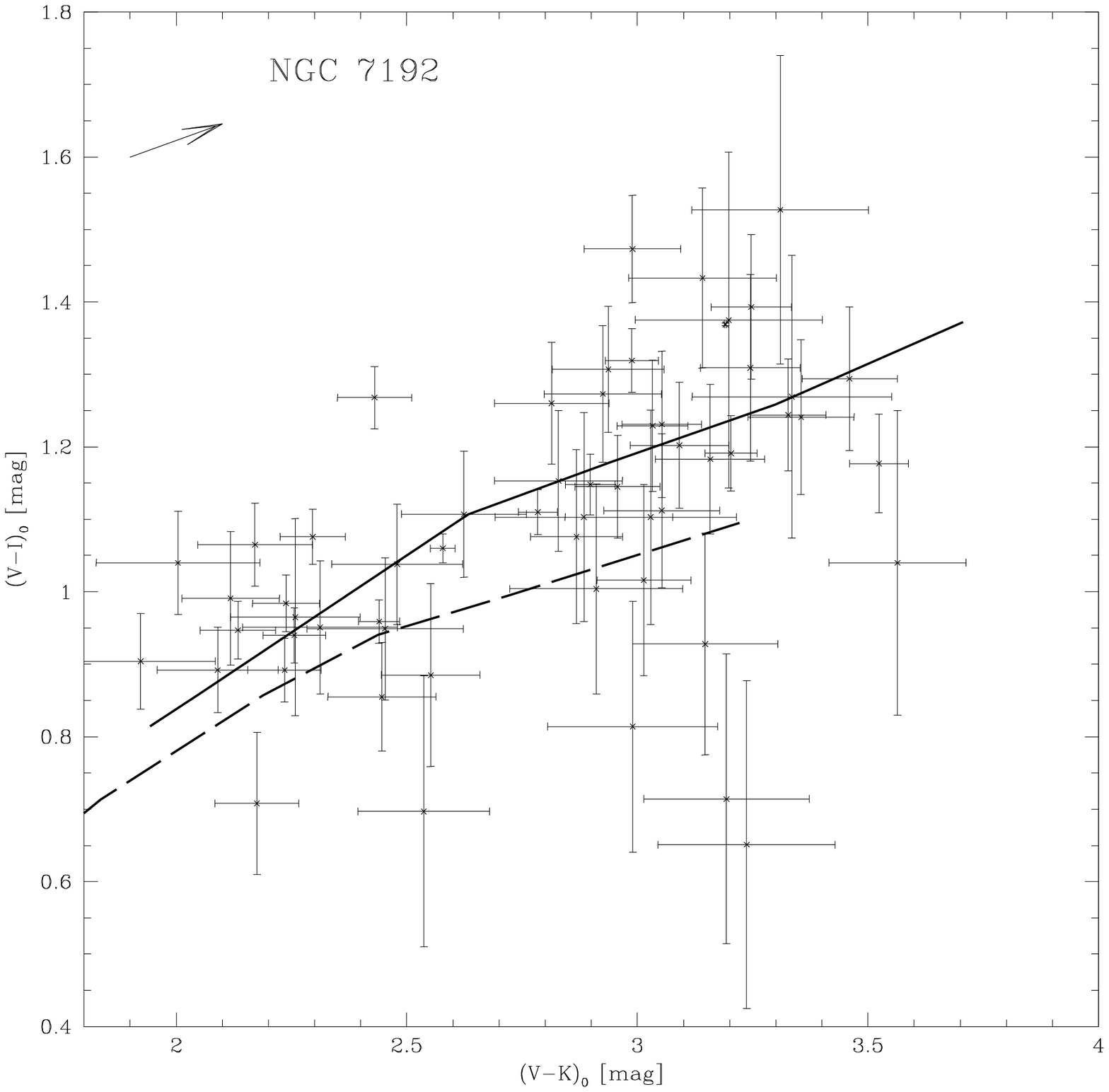}
\caption{Colour-colour diagram for NGC~7192. No clear indication for
a second red sub-population can be found (isochrones as in Fig.\ref{f:colcoln5846}). }
\label{f:colcoln7192}
\end{figure*}

Referring to the CMD and the colour-colour diagram for NGC~7192 given
in Figure \ref{f:n7192cmd} and Figure \ref{f:colcoln7192}, the
difference to the globular cluster system of NGC~5846 can be 
seen: while NGC~5846 exhibits a strong spread 
perpendicular to the isochrones, NGC~7192 is more homogeneously populated
along the isochrone. 

The small number of objects do not allow one to draw strong
conclusions from the KMM test in terms of the existence of colour
bimodality. The colour-colour plot for NGC~7192 does suggest the
existence of two populations of different metallicity (i.e., a
`metal-poor' population with $(V-I)<1.1$ and $(V-K)<2.8$ and a
`metal-rich' population with $(V-I)>1.0$ and $(V-K)>2.8$) but the
small number statistics do not allow one to support this firmly
statistically.

Formally, the colour for the two peaks of the distribution projected
on the $V-K$ axis is returned by KMM to be $(V-K)$= 2.83 and 2.89
(with 90\% of the objects assigned to the first peak), if a bimodal
distribution is assumed. Since the difference in colour between the
two peaks is well below the photometric error it seems far-fetched to
assume the existence of two distinct globular cluster sub-populations.

Therefore we assume that the NGC~7192 globular cluster system consists only of one dominant 
population. Following the calibration values by Kissler-Patig (Paper
I) we derive a peak metallicity of [Fe/H]=$-0.57\pm0.37$. 

\section{Determining the age structure from colour-colour diagrams}
\begin{table*}[]
\centering
\caption{Fit parameter for model isochrones (\cite{bruz00}) for a
logarithmic fit $(V-I)$=~A$\cdot$log$(V-K)$+B.}
\label{tab:parameter}
\begin{tabular}{lcccccccc}
\hline
\noalign{\smallskip}
Isochrone [Gyr] & 1 & 2 & 3 & 5 & 7 & 10 & 13 & 15 \\
\noalign{\smallskip}
\hline
\noalign{\smallskip}
A & 0.5645 & 0.7136 & 0.7674 & 0.7795 & 0.7922 & 0.8134 & 0.8423 & 0.8573 \\

\noalign{\smallskip}
\hline
\noalign{\smallskip}
B & 0.3296 & 0.2809 & 0.2633 & 0.2765 & 0.2826 & 0.2766 & 0.2596 & 0.2502  \\
\noalign{\smallskip}
\hline
\end{tabular}
\end{table*}

\subsection{Colour distributions}
\label{s:model}
The basic idea in using optical and near-infrared photometry in
globular cluster studies is to lift the age -- metallicity degeneracy
and to resolve possible age sub-populations of globular clusters. The
minor drawback with respect to using optical colors alone is an increased 
photometric error mainly caused by the infrared observations, i.e., in 
the K-band.  This, however, is more than compensated for by the larger
diagnostic power of the optical -- near-infrared combination (see Papers I
and II). A separation in the colour-colour plot of sub-populations with
different metallicities becomes easier, and a separation in age at a
given metallicity becomes feasible. 

We investigate below what the diagnostic power of the method is in terms
of separating sub-populations of different ages.
To do so, we investigated artificial colour-colour distributions of composite
populations based on the Bruzual \& Charlot (2000) isochrones. These
artificial systems were built as follows:

$\bullet$ {\bf In a first step} we create an artificial $V-K$ distribution of 
metal-rich globular clusters. We populate the red $(V-K)$ colour range 
($2.7 \leq (V-K)\leq 3.8$), assuming that it is occupied by {\bf{``old''}} and
{\bf{``intermediate''}} age objects. We do not consider bluer clusters
($2.0\leq(V-K)\leq 2.7$), assuming that these are only {\bf{``old''}} objects. 

For now, we assume cases similar to NGC~5846 (see Figure
\ref{f:colcoln5846}) in terms of sample size and photometric errors.
For a first exemplary case, we assume 50$\%$ of the red population to
be 15 Gyr old and 50$\%$ to be 3 Gyr old. These numbers are not
assumed to reflect the `true' situation, but rather serve to
demonstrate the method at this point. We will probe models with
different ratios in the future (see Section \ref{s:future}). Further,
it will become clear below that we are not directly comparing observed
and simulated colour-colour diagrams but rather their cumulative age
distributions, i.e.,~properties of the distributions still need to be
``calibrated''.

The final modeled systems contain 43 old, blue objects (not considered 
further) and 120 red objects homogeneously distributed within the $V-K$ range 
\footnote{We experimented also with Gaussian distributions within this colour 
range, but the final results of the experiment did not differ
significantly and we have no better physical justification for a gaussian than for a
homogeneous colour distribution}. 
The red population was divided into an old (15 Gyr) and young (3 Gyr)
population with 60 objects each.  

$\bullet$ {\bf In a second step} we attach to each $V-K$ data point a
($1 \sigma$) error drawn randomly from our observed list of $\Delta(V-K)$ 
errors for NGC~5846, and then smear in a Monte-Carlo approach each $V-K$ point 
with up to $\pm 3$ times its associated error (i.e.~allowing up to $3
\sigma$ errors in very rare cases). These new $V-K$ values are stored
with their 1 $\sigma$ error and used for the further process.

\begin{figure}[!]
\centering
\includegraphics[bb=50 150 580 710, width=8cm]{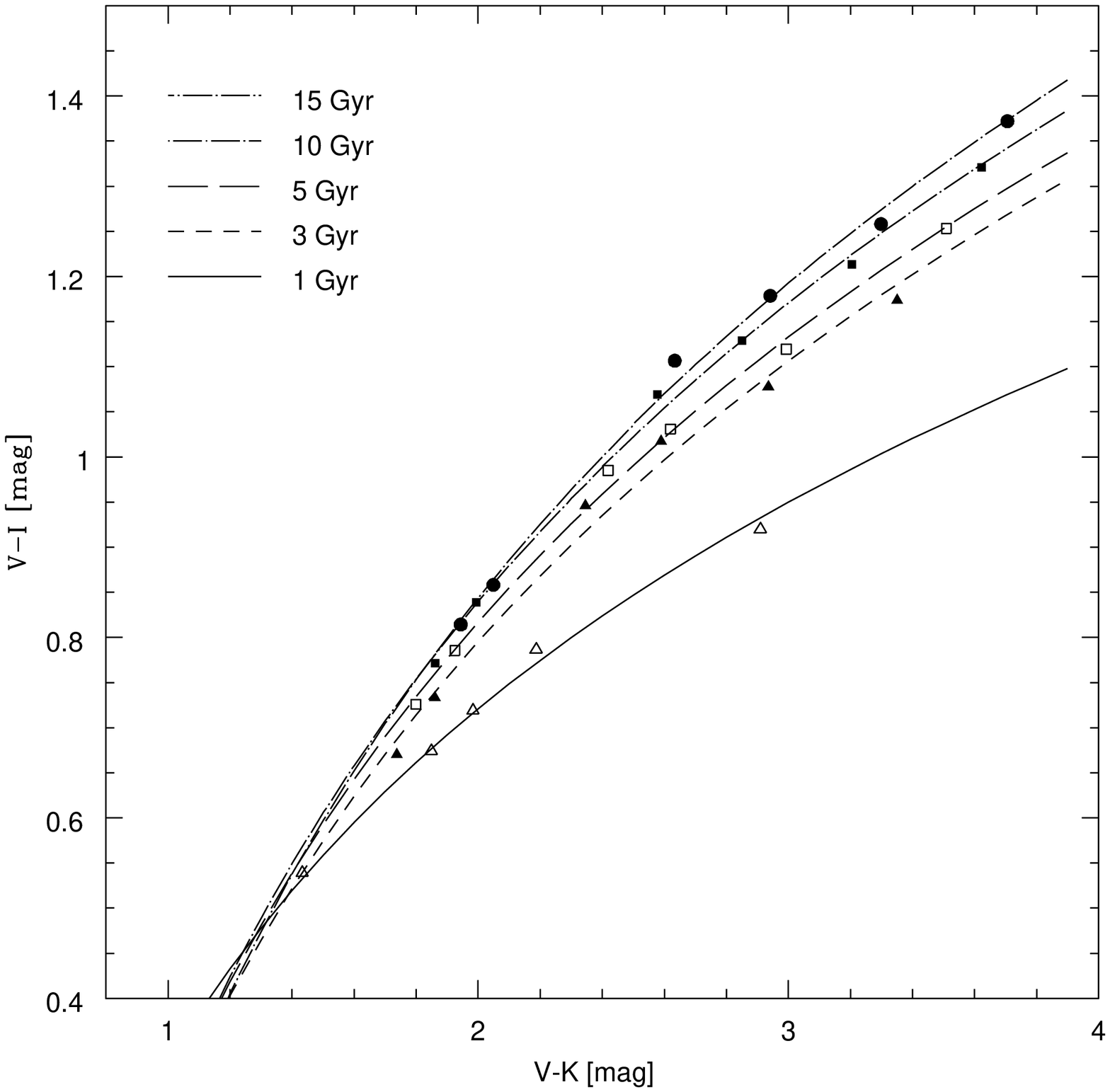}
\caption{SSP isochrone fit for $(V-I)$ vs.\ $(V-K)$ colour-colour diagrams. The
 symbols are colours given by the SSP models (\cite{bruz00}) for 1 Gyr
 (open triangle), 3 Gyr (solid triangle), 5 Gyr (open square), 10 Gyr
 (solid square) and 15 Gyr (dots). The lines represent the result of a
 logarithmic fit to the isochrones with
 $(V-I)$=~A$\cdot$log$(V-K)+B$. The fit parameters A and B are given
 in Table 2.}
\label{f:bcmodel}
\end{figure}

$\bullet$ {\bf The third step} consists of associating a $V-I$ colour
to each new $V-K$ colour. To do this, we perform a least-square fit to
the SSP model isochrones (in this case from Bruzual \& Charlot 2000)
by a logarithmic function ($(V-I)$=A$\cdot$log$(V-K)$+B). The
particular fit parameters A and B are given in Table 2. Figure
\ref{f:bcmodel} shows the isochrones as given by \cite{bruz00} and our
fits to the isochrones (solid lines). 

The fits are then used to compute for each $V-K$ point the
corresponding $V-I$ point, once the age was chosen. In our case, we used
the 15 Gyr fit for the 60 old artificial clusters, and the 3 Gyr fit for the
other 60 (young) artificial clusters.\\

$\bullet$ {\it Finally, in \bf the fourth step} we associate a measurement
error $\Delta(V-I)$ to each $(V-I)$ data point in a similar way as for
$(V-K)$. We now have 
a set of 120 objects (60 old, 60 young) with associated $(V-K)$ and
$(V-I)$ colors and errors.\\

\begin{figure}[!]
\centering
\includegraphics[bb=50 150 580 710,width=8cm]{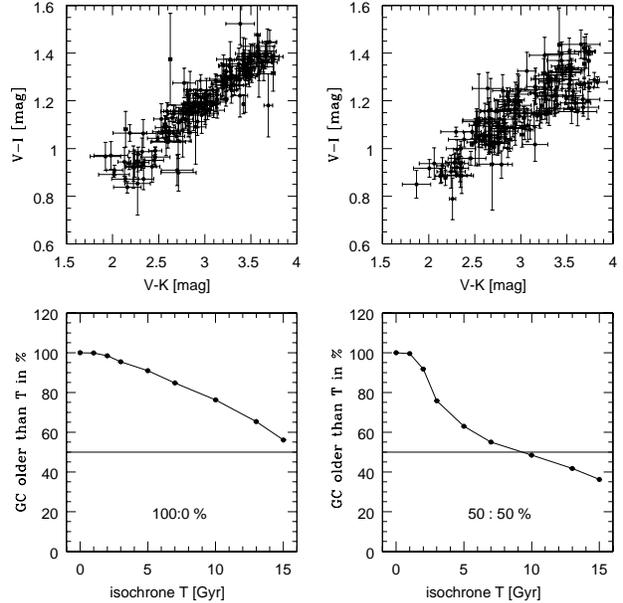}
\caption{The upper two panels show the $(V-I)~$vs.$~(V-K)$ colour-colour
diagram for a single age population (left panel) and a composite 50\%
old (15 Gyr) and 50\% young (3 Gyr) cluster population (right panel).
In the lower panels, the cumulative age structure (see \S \ref{s:agedis}) 
for both cases is given as the mean of 1000 such simulations. The 50$\%$-level is marked as a solid line. }
\label{f:colcolmodel}
\end{figure}

As an example of such artificial colour-colour distributions, we show in
Figure \ref{f:colcolmodel} the modeled colour-colour diagram of a
purely 15 Gyr old population and a composite {\bf old} (15~Gyr)
and {\bf young} (3~Gyr) clusters as well as the resulting age
distribution (lower panels). This distributions should be compared
to the observed data for NGC~5846 in Fig.~\ref{f:colcoln5846}.

\begin{figure}[!ht]
\centering
\includegraphics[bb=50 160 580 690, width=8cm]{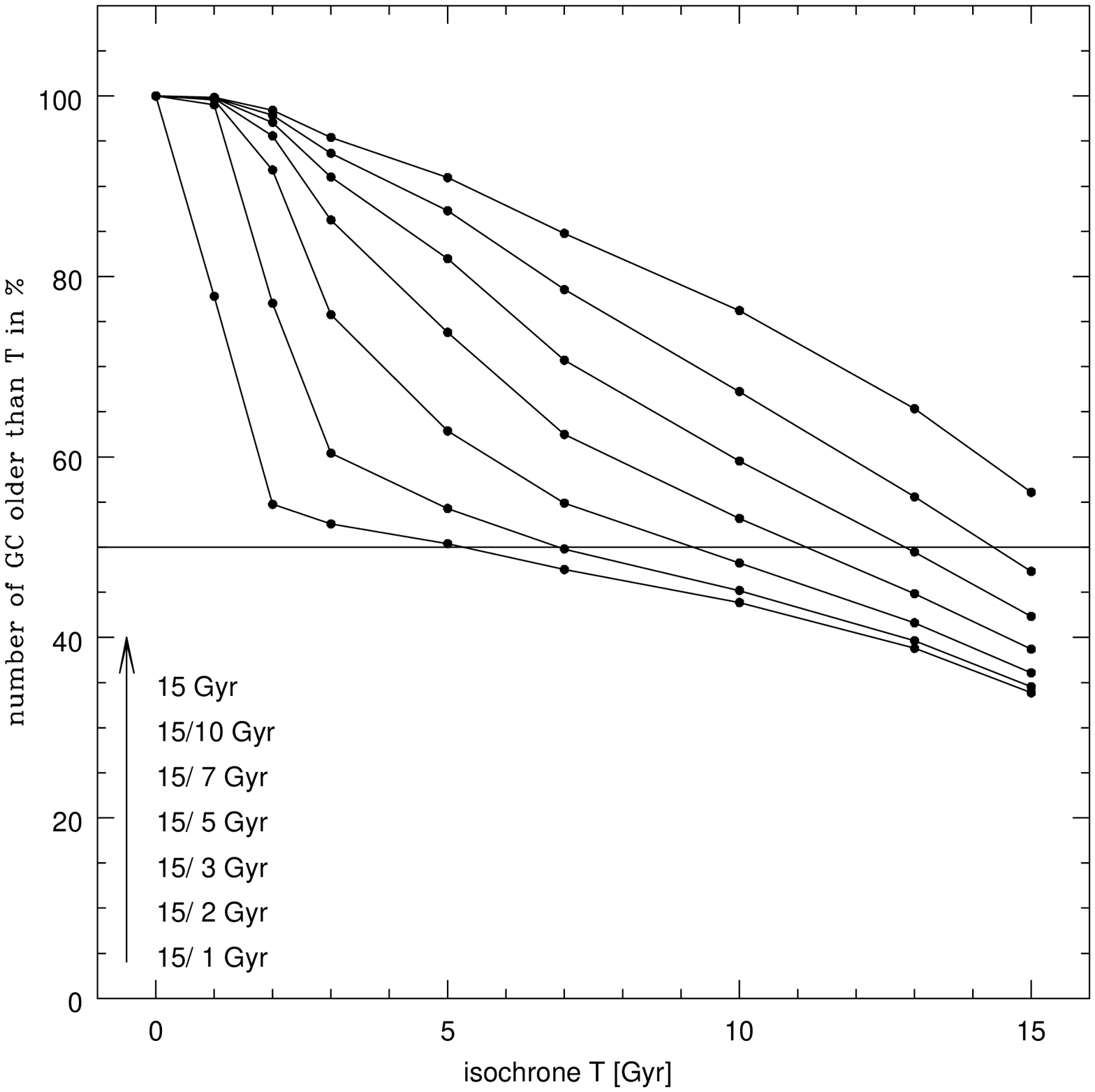}
\caption{Age distribution for mixed cluster populations assuming half
of the objects being 15 Gyr old and half of them being respectively 
1, 2, 3, 5, 7 or 10 Gyr old. The 50$\%$ level is marked by a solid line.}
\label{f:agemodel}
\end{figure}

\subsection{Cumulative age structure}
\label{s:agedis}

The separation of the isochrones in Fig.~\ref{f:bcmodel} and the difference
between the distributions in Fig.~\ref{f:colcolmodel} (upper panels) motivated 
us to use the optical-near infrared photometry in order to try to separate 
globular cluster populations of different ages. This appears from the
above plots to be at least possible to age differences of 10 Gyr or more
(with respect to 15 Gyr for the oldest clusters). 
Therefore we decided to take the analysis one step further
and to make a first attempt towards determining the age structure of the 
globular cluster systems. 

We first attempted to compare directly the observed and the artificial
colour-colour distributions, but this turned out to be too complicated. Two
dimensional statistical tests are required in that case and the typical
number of data points is not leading to any statistically significant
results. We experimented with 2-dimensional f-test, 2-dimensional
Kolmogorov-Smirnov tests, but in order to constrain the tests somewhat, one
needs to put rather artificial constraints and the physical meaning of the
final results is  dubious.

We therefore decided to focus on 1-dimensional representations of the
age structure. The dimension was chosen to maximize the age gradient
along it, i.e.,~``perpendicular'' to the isochrones.

\begin{figure*}[!ht]
\centering
\includegraphics[width=8cm]{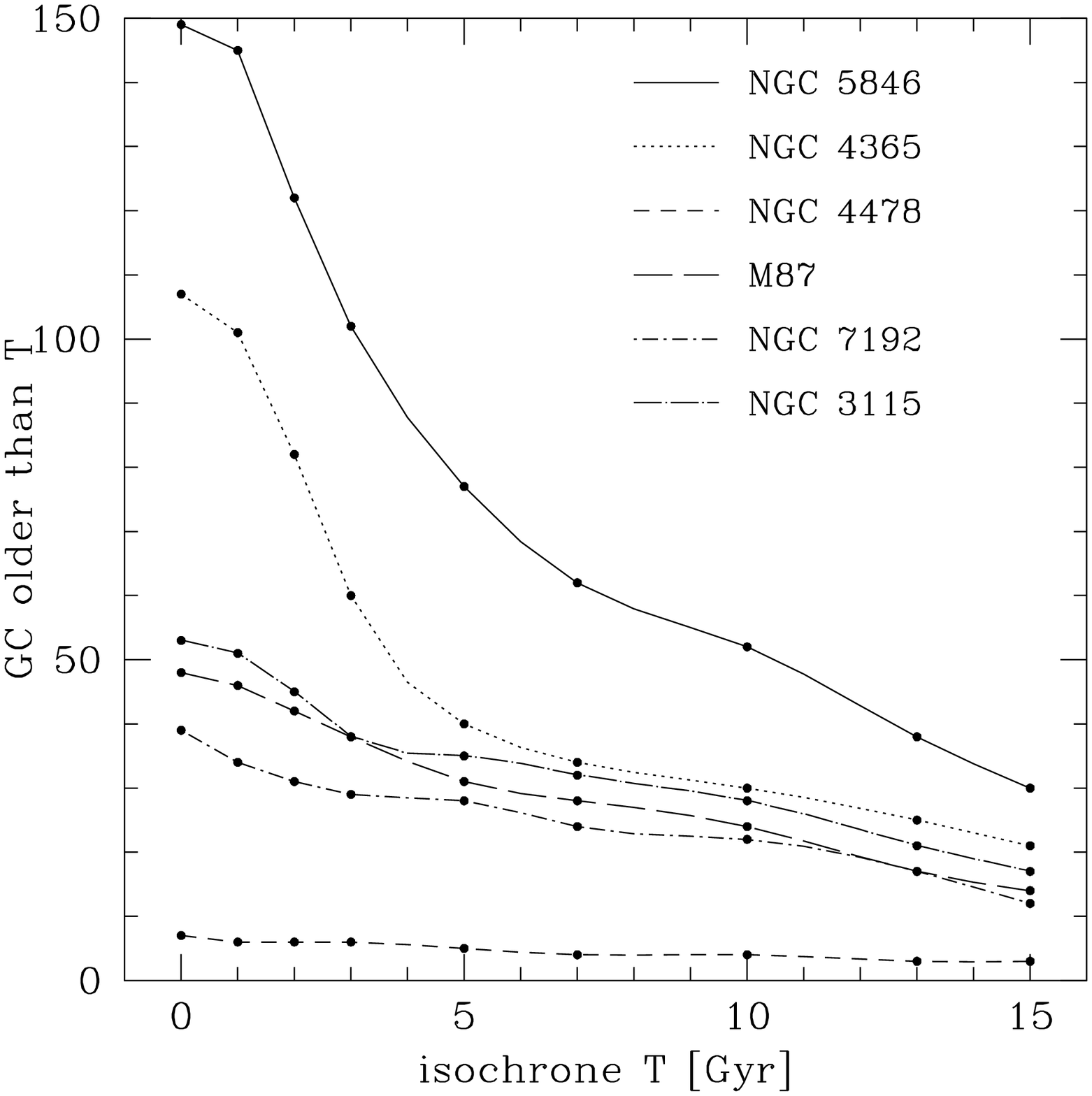}
\includegraphics[width=8cm]{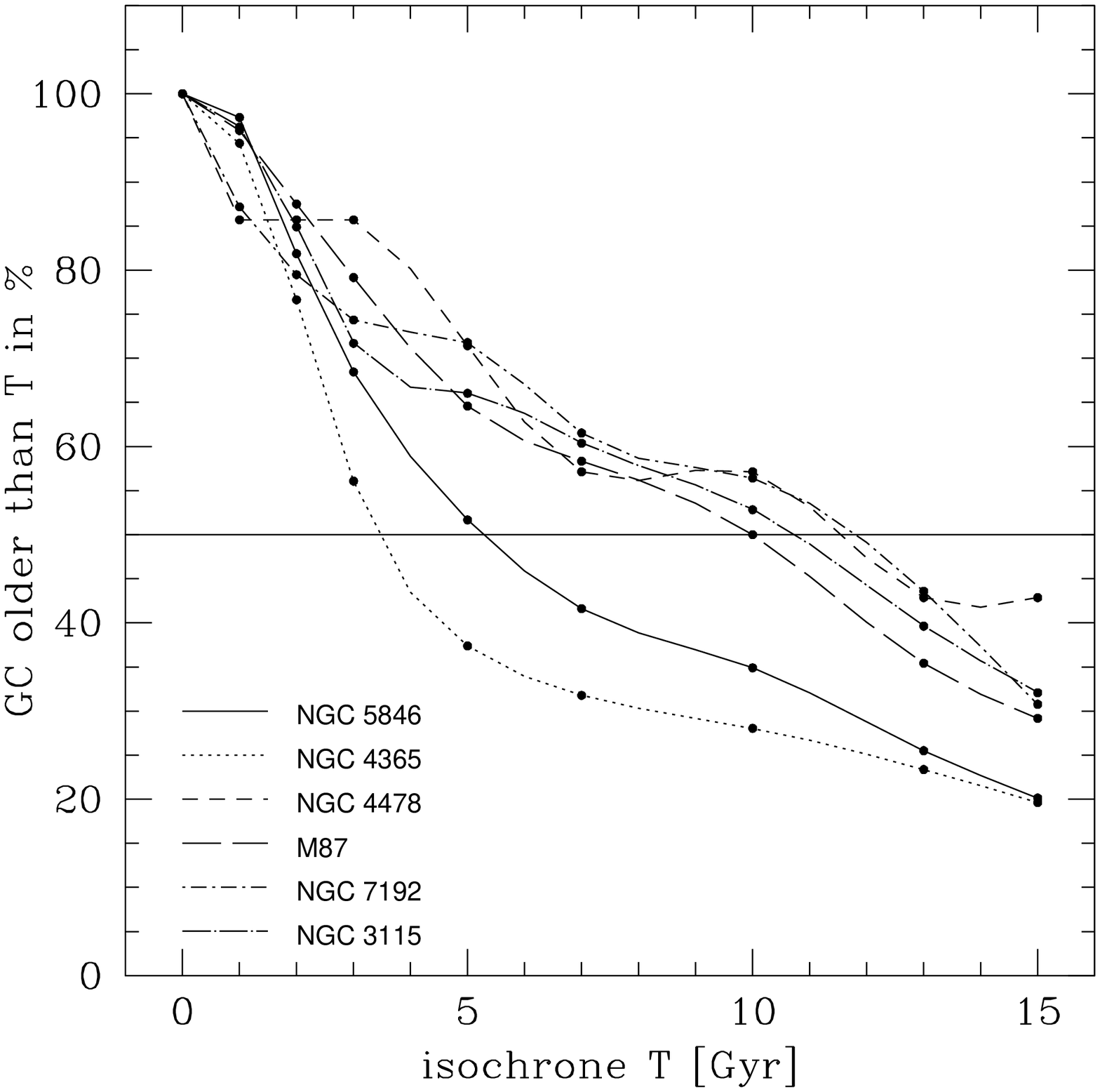}
\caption{Age distribution for different galaxies (see text). 
The absolute number counts (left panel) are normalised to the total numbers 
of clusters in the sample (right panel). It is clearly seen that
two systems (NGC~5846, NGC~4365) are significantly different from
NGC~4478, M~87, NGC~7192 and NGC~3115. This is interpreted as NGC~5846
and NGC~4365 hosting intermediate-age sub-population (which was
confirmed spectroscopically for NGC~4365). The 50$\%$ level is marked by a
solid line.}
\label{f:galaxages}
\end{figure*}

Briefly, we associate to every cluster in our observed or artificial
distribution an ``age greater than X'' when it lies above the
isochrone of that age X in the colour-colour diagram. We start with
the youngest isochrone (0 or 3 Gyr, see below) for which most cluster
will lie above, and then move up isochrone by isochrone (in the steps
1,2,3,5,7,10,13,15 Gyr). Hereby the notation ``0 Gyr'' refers to
objects below the 1 Gyr isochrone. The result is an inverted
cumulative distributions as shown in Fig.~\ref{f:agemodel} for
artificial distributions and in Fig.~\ref{f:galaxages} for the
observed distributions of all our galaxies analyzed so far. The
cumulative age distribution can be represented in absolute numbers
(left panel) or normalized to the total number of objects (at an
arbitrary bin, right panel).

\subsubsection{Artificial data sets}

As described above, the first set of simulations was done for combinations of
a 15 Gyr old sub-population and an equal number of intermediate-age
objects (1,2,3,5,7,10 Gyr). The results are shown in Figure \ref{f:agemodel}. 
Each curve shows the mean age structure as evaluated from 1000 models for 
that given age composition. 

The realistic photometric errors create a spread in the colour-colour
diagram such that even a pure 15 Gyr system does not show 100\% of the
clusters to be older than 15 Gyr. Instead, the spread in the $(V-I)$
vs.~$(V-K)$ diagram leads to a gentle fall-off with isochrone age.

However, this fall-off clearly changes when a second, younger
sub-population gets mixed in. By the time one mixes a 1,2,3 Gyr
population in the system, the fall-off becomes very steep around 2,3,4
Gyr, and the curves cross the 50\% level well before the 10 Gyr mark.

\subsubsection{Observed data sets}

The result for the artificial distributions can be compared with the observed 
age distributions of the globular cluster systems 
(see Figure \ref{f:galaxages}) in galaxies analyzed so far (M~87, NGC~4478, 
NGC~4365, NGC~3115, NGC~5846 and NGC~7192). 

Taking first the observed data alone, we notice a clear similarity between
NGC~5846 and NGC~4365, as opposed to the 4 other systems (see right panel
of Fig.~\ref{f:galaxages}. Both galaxies fall-off steeply at early
isochrone age and cross the 50\% line well before 10 Gyr.

When compared to the results of artificial distributions, this leads
immediately to the interpretation that both must host a significant
fraction of intermediate-age clusters within their red sub-population.
For NGC~4365, this was suspected already in Paper II and has since then
been confirmed spectroscopically (Larsen et al.~2002).

In contrast, the age distribution of NGC~7192, NGC~3115,
NGC~4478 and M~87 seems to be more consistent with what we would expect
for a single age and old ($\geq$10 Gyr) population.

\subsection{Contamination of background objects}
\label{contamination}

Contamination of background objects is a potential problem in our
analysis and we briefly investigate its impact below.

As above for the artificial distributions, we adopt a situation similar
to the observations of NGC~5846. 
We used the Hubble~Deep~Field South (available at:
www.stecf.org/hstprogrammes/ISAAC), which presents two advantages: it
covers exactly the same (WFPC2) field of view as our observation, and it has 
a deep enough $K$-band observation to match our ISAAC observations of the
globular cluster systems ($K<21.5$ mag). The HDF-S sample was further
selected in colours as for our samples. However, we could {\it not} apply a
FWHM selection  
(having only the list of objects) so that the contamination is expected to be 
an overestimate. Even in that case, however, we show below that the effect is 
negligible for cases such as NGC~5846.

\begin{figure}[!ht]
\centering
\includegraphics[bb=50 420 580 710,width=9cm]{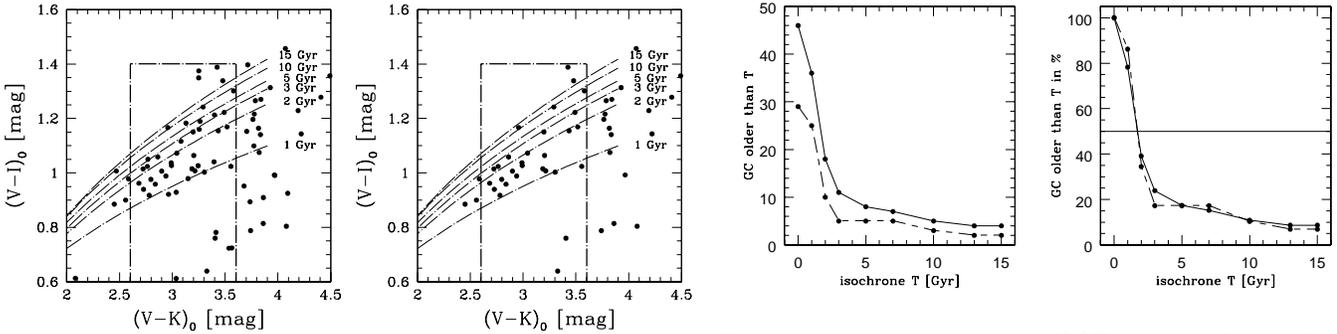}
\caption{$(V-I)$~vs.~$(V-K)$ colour-colour diagram for the
Hubble~Deep~Field-South using different $K$ completeness limits: left,
all objects with $K<21.5$, right all objects with $K<21.0$ (bracketing
our completeness limit for NC~5846).  The box
marks the colour range in $(V-I)$ and $(V-K)$ used for the
determination of the globular cluster age structure as described in section
\ref{s:agedis}. Both diagrams show that the highest contamination of 
our sample is expected below the 2 Gyr isochrone. The isochrones
superimposed are from Bruzual \& Charlot SSP models (2000).}
\label{f:hdf}
\end{figure}

\begin{figure}[!h]
\centering
\includegraphics[bb=50 420 580 710,width=8cm]{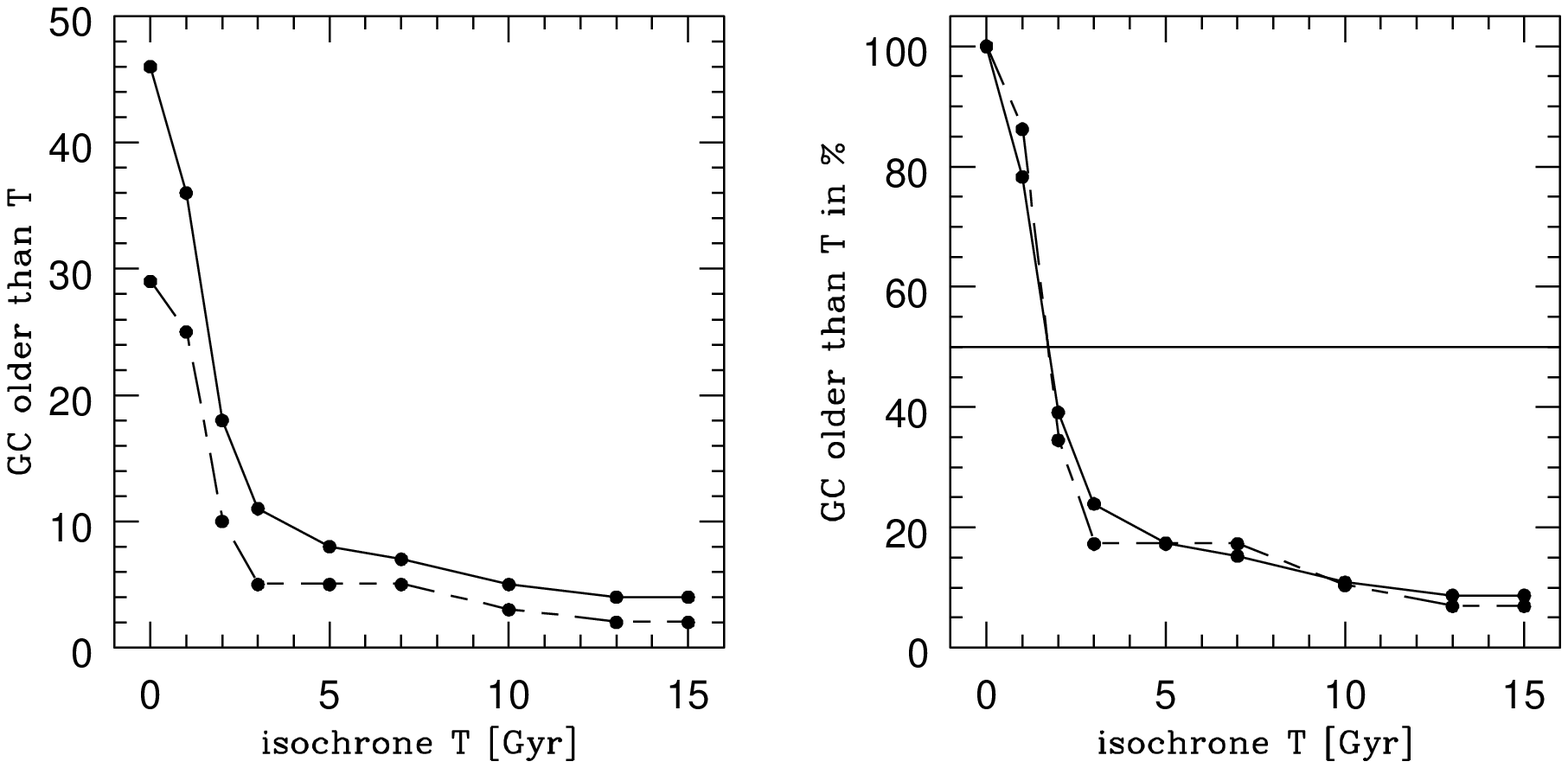}
\caption{Age structure of the HDF-South objects using the procedure
given in Section \ref{s:agedis}. Left and right panels show the absolute
and relative age distributions respectively.
The colour selected samples with $K<21.5$ and
$K<21.0$ are shown as solid and dashed curves, respectively.
The 50$\%$ level is marked by a solid line.}
\label{f:hdfagedis}
\end{figure}

Figure \ref{f:hdf} shows the colour-colour diagram of the HDF-S.
Depending on magnitude selection, we have between 25 and 40 background
objects in our colour selection box. The resulting cumulative age
distribution for the HDF-S sample is shown in Figure \ref{f:hdfagedis}
and assigns about 50$\%$ of the objects to a population younger than 3
Gyr. Further, from the colour-colour plots, it becomes clear that the
majority of these objects actually lie below the 2 Gyr line, being
bluer in $(V-K)$ and $(V-I)$ than the intermediate-age, metal-rich
globular clusters in NGC~5846. It is most probably dominated by a
star-forming galaxy population that would be rejected by our FWHM
criteria.

\begin{figure}[!h]
\centering
\includegraphics[bb=50 420 580 710,width=8 cm]{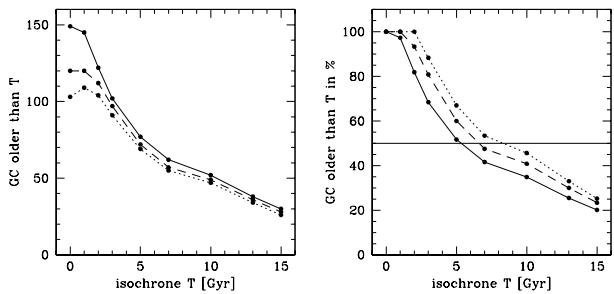}
\caption{Influence of background contamination on the age structure of the 
system in NGC~5846. The left and right panels show the absolute and relative 
cumulative age distributions, respectively.
The uncorrected distribution is shown as a solid curve. The distribution
corrected for background contamination of $K<21.5$ and $K<21.0$ are shown as 
long dashed and short dashed lines, respectively.}
\label{f:n5846corr}
\end{figure}

The effect of contamination on NGC~5846 is shown in Figure
\ref{f:n5846corr}.  There, we plot the uncorrected (absolute and
relative) age distributions, as well as the ones corrected for
background contamination using the HDF-S source counts. Clearly, the
effect is marginal. The contaminating objects tend to drag the age
distribution towards younger ages, mimicking a slightly younger
sub-population, e.g.~the intersection with the 50$\%$-level occurs at
a larger age. Thus, ``old'' distribution such as for M~87, NGC~3115
etc would appear even older when corrected for contamination. In the
cases of NGC~4365 and NGC~5846 (large numbers of clusters) the effect
of this (overestimated) contamination is small and does not influence
the conclusion that these systems host a significant intermediate-age
population.

\section{Conclusions and future work}
\label{s:discu}

\subsection{Conclusions}

In this paper, we presented results of our optical\,--\,near-infrared study of the globular cluster
systems of NGC~5846 and NGC~7192. 
While the latter does not show any significant anomaly, 
the former shows a case very similar to that of NGC~4365
(see Paper II) with good evidence for a significant metal-rich, 
intermediate-age population in addition to an old, metal-rich
one.
 
We developed a new method to quantify the age structure of the
globular cluster systems studied so far, based on a comparison of the
cumulative age distributions of the observed and modeled
optical\,--\,near-infrared colour-colour diagrams.  This method
appears powerful enough to detect intermediate-age sub-population
within globular cluster systems.

Our first conclusions for the galaxies studied so far are that
both NGC~4365 and NGC~5846 have a cumulative age distribution of
the metal-rich globular clusters that is different from those of the 4 other
galaxies (NGC~3115, NGC~4478, NGC~7192 and M~87). The former are better
modeled by  
by a composite metal-rich population including an old (15 Gyr) as well as a 
significant young (1-5 Gyr) population. The others, in
contrast, are very similar to the model for pure old populations and are
thus best explained by being dominated by old objects.

NGC~7192 suffers from small number statistics, but the current dataset is,
with respect to the colour-colour diagram and the age structure, more
consistent with only one sub-population. Compared to the globular
cluster systems investigated so far, NGC~7192 most closely resembles 
NGC~3115. 

For M~87 and NGC~4478 the results given in Paper I were not fully
conclusive. NGC~4478 also suffers from small number statistics. M~87
appears now more clearly dominated by two sub-populations differing in
metallicity, but with a metal-rich population dominated by old objects
(see also Jord\'an et al.~2002). \\

We need to emphasize again that the method cannot, currently, produce
reliable {\bf absolute} numbers, neither in terms of ages, nor in terms
of ratios between the different populations. The results in this paper
should therefore be considered as qualitative for now, and will be
better quantified in the future.


\subsection{Future work on age dating}
\label{s:future}

The upcoming papers will analyze the remaining galaxies in our dataset 
(in total 11 galaxies). This will allow us to discuss the results in the light 
of galaxy properties.
In particular, we will look at trends with galaxy size and environment.
The local density (\cite{tully88}) covered by our  galaxy sample spans from
0.08 Mpc$^{-3}$ (NGC~3115) to 4.17 Mpc$^{-3}$ (M87) and environmental effects 
appear to be an important ingredient to galaxy formation and
evolution.\\

On the modeling side, our goal is to improve the quantitative
information of our method.
We are developing a $\chi^2$ test to find the best solution in the two 
parameter space of the models: ratio old/young and age of the young population,
which are slightly degenerate. We also plan, with the help of spectroscopy and 
wide field photometry, to be able to better calibrate the models in terms
of absolute age.

Also, we are exploring the dependence of the results on a particular
SSP model and we will repeat the determination of the age structure using
SSP models by Maraston (2000) and Vazdekis (1999). 
Further, we will investigate in more detail the effect of background
contamination. This latter aspects will be the subject of a separate
paper.


\begin{acknowledgements}

The authors would like to thank the ESO user support group and the ESO
science operations for having carried out the program in service
mode.  We are also grateful to Stephane Charlot for providing his
population synthesis models prior to publication. M.~Hilker
acknowledges support through Proyecto Fondecyt 3980032. THP gratefully
acknowledges the support by the German \emph{Deut\-sche
For\-schungs\-gemein\-schaft, DFG\/} project number Be~1091/10--2.  DM
is supported by FONDAP 15010003 Center for Astrophysics.
\end{acknowledgements}


\end{document}